\documentclass[doublecol]{epl2} 

\title{Multi--hump soliton--like structures in interactions of lasers
  and Bose--Einstein condensates}
\shorttitle{Multi--hump structures in laser--BEC interactions} 

\author{F. Cattani\inst{1} \and A. Kim\inst{2}  \and T. Hansson\inst{3} \and
  D. Anderson\inst{3} \and M. Lisak\inst{3}}
\shortauthor{F. Cattani \etal}

\institute{                    
  \inst{1} School of Mathematics, University of Southampton,
  Southampton, SO17 1BJ, UK\\
  \inst{2} Institute of Applied Physics, Russian Academy of Sciences - 603950 Nizhny Novgorod, Russia\\
\inst{3} Department of Radio and Space Science, Chalmers
University of Technology - SE-412 96 G\"{o}teborg, Sweden
}

\abstract{An investigation is made of multi–hump and periodic solutions of the semi–classical
coupled equations describing laser radiation copropagating with a Bose-Einstein condensate. Solutions
reminiscent of optical vector solitons have been found and have been used to gain understanding
of the dynamics observed in the numerical simulations, in particular to shed light on the
phenomenon of jet emission from a condensate interacting with a laser.
}

\begin{document}

\maketitle

\section{Introduction}

The experimental realisation of Bose-Einstein condensates (BEC) and of
coherent atomic beams has inspired a renewed effort in the theoretical
modelling of such physical systems. Of particular interest from the
point of view of applications, is the  possibility of manipulating
atomic beam structures by using their interactions with light. Besides, due to
the mathematical modelling of these interactions, these phenomena can
also offer a possible test of the analogies between optics and quantum
matter waves, which is what we would like to explore further in this
work.\\ 
It
is well known that the dynamics of a BEC satisfies an 
equation which has the same mathematical structure of the nonlinear
Schr\"{o}dinger equation, the fundamental equation of nonlinear
optics, \cite{ref:review}, the Kerr--like nonlinearity being due to
the atom--atom interactions which can be either focusing or defocusing
depending on the sign of the scattering length. Considering the interactions of a BEC with
laser radiation brings new terms into the atom equation whose form depends
on the modelling of the interactions, the simplest form being the
dipole--dipole one. Saffman started to study the
consequences of these interactions, \cite{ref:models} while 
Krutistky {\em et al.} derived a fundamental model from first
principles within the framework of quantum field theory and later
generalized it to consider atom transitions (in the case of high atom densities),
\cite{ref:krutitsky}. More recently, the same structure for the atom
equation was derived within a semi--classical approach,
\cite{ref:us3}. The idea there was to work out an expression for the
force exerted by light on atoms from basic classical physics and use
it as a potential term in the  Schr\"{o}dinger equation of the atoms. 
It is important to notice that, within the limits of a semi--classical approach,
the atom equation agrees with the more general equation of \cite{ref:krutitsky}. The point
is that a natural consequence of BEC atoms interacting with
photons is the emergence of a coupling between the dynamics of the
atoms and that of the laser radiation. In fact, what these models
describe is photons from the incident radiation exciting atoms which in turn re--emit
photons, which are then absorbed by other atoms, thus giving rise to a long-range
interatomic interaction, \cite{ref:krutitsky, ref:cohen}. This obviously has an
effect not only on the atoms but also on the radiation. This gives rise to a coupling term in the atom
equation and the need for an additional equation describing
also the dynamics of the radiation.  It is this
coupling that can bring about novel effects. 
It has been shown via numerical simulation of the coupled dynamics
that self-localised structures and mutual atom--light guiding can be
achieved exploiting the dipole--dipole interactions, \cite{ref:us1}.
In particular it was
observed that solitary--like localised
structures can be generated and emitted from a central bunch of atoms, \cite{ref:us2}.
This is suggestive of analogous optical effects such as soliton
ejection, see \cite{ref:emission} and references therein,  and we will try here to obtain an understanding of how the structures
ejected in laser--BEC interactions are created and emitted. The
peculiarity of the emission discussed here is that the generation of
the solitary--like structures  is entirely due
to the atom--laser coupling and their ejection from the region where
they have been created can be ascribed to their tail--tail
interaction. This sort of interaction, being the structures initially well
separated, is determined by the overlap of the structures themselves,
{\em i.e.} the cross terms arising when calculating terms such as
$|\psi_{TOT}|^2$ if $\psi_{TOT}$ can be written approximately as the sum of two or
more localized functions. Since these functions are assumed to be
spatially separated, the only contributions to the interaction term
$|\psi_{TOT}|^2$ will come from their overlapping tails.\\
In optics, for example in the early study  of Gubbels
{\em et al.}, \cite{ref:gubbels}, up to the more recent work of
Assanto {\em et al.}, \cite{ref:assanto}, the emission is engineered
in such a way that the trap or guide does not undergo any
modification. Similarly in studies of soliton emission from BECs, see
for example \cite{ref:rodasverde}, the basic physical mechanism for
emission concerns the atom--atom interactions and the trap is not
modified. In our case, the atoms modify the medium through which they
propagate, i.e. the laser radiation, which in turn affects the atom
propagation via a nonlinear coupling. If mutual trapping can be
achieved, localised symbiotic 
atom--light structures will start to propagate together and the
emission will be due to the interaction of these newly born structures
rather than 
to tunneling or engineering of the trap.  This process is 
also  reminiscent of the solitonic gluons studied by Ostrovskaya {\em et al.}, \cite{ref:elena}.\\
Inspired
by numerical simulations which clearly show how, during the coupled
propagation of laser and atoms, symmetric and mutually localised structures are
formed and then ejected, the first step we would like to take is an
investigation into the possible existence of multi--hump stationary
solutions of the coupled equations. This kind of solution is not found in
the usual nonlinear Schr\"{o}dinger equation but is known to be a
possibility in coupled systems of nonlinear equations such as those
describing birefringent fibers. 
 If multi--hump solutions  exist
for the laser--BEC system as well, it should be possible to use
known methods of nonlinear optics and to study the
interaction between the different soliton--like formations to
infer whether there will be mutual repulsion or attraction or whether
they can propagate together. This would justify the numerical results
and provide an insight into the possibility of using these effects to
manipulate localised atom--light structures.\\
In what follows we will briefly review the basic coupled model and
stationary coupled equations as well as
the limitations imposed by a semi--classical description. We will then
search for multi--hump solutions and study in particular  the tail interaction of
double--peaked solutions to
show how, within
this model,  it is possible to expel solitary--like structures in a
jet-like fashon, or even
having them colliding and merging together into a single central structure.

\section{Model equations}
The semi--classical coupled stationary equations as introduced in \cite{ref:us1,ref:us3}
are given by
\begin{eqnarray}\label{eq:atom}
   \hbar\omega_a \Phi =
   \hat{H}_0\Phi
   +\left[U_0\Phi^2-
   \frac{\alpha}{4}\frac{{\boldsymbol{\mathcal{E}}}^2}{\left(1-\frac{4\pi}{3}\alpha
   \Phi^2\right)^2}\right]\Phi,\\
\nabla^2 {\boldsymbol{\mathcal{E}}} + k_L^2 \left(1 + \frac{4\pi\alpha
    \Phi^2}{1-\frac{4\pi}{3}\alpha\Phi^2}\right){\boldsymbol{\mathcal{E}}}=0.\label{eq:laser}
\end{eqnarray}
where for the full atom wave function and the full laser field it has
been assumed
\begin{eqnarray}
\Psi({\bf{r}}, t) = \Phi({\bf{r}}) \exp(-i \omega_0 t),\\
{\bf{E}}({\bf{r}}, t) =   Re[{\boldsymbol{\mathcal{E}}}({\bf{r}}) \exp(-i \omega_L t)].\\
\end{eqnarray}
Here,  $\alpha(\omega)=-d^2/\hbar\Delta$ is the atomic  polarizability at the
laser frequency $\omega_L= k_L c$, with $\Delta=\omega_L -\omega_a$ being the
detuning from the nearest atomic resonance frequency $\omega_a$, and
$d$ is the dipole matrix element of the resonant
transition. Besides,  for the gas density we have $n=|\Psi|^2$, $U_0=4\pi\hbar^2a_s/m$, $m$
is the atom mass and $a_s$ is the $s$--wave scattering length
(which will be assumed positive as for repulsive atom-atom
interactions) and $\hat{H}_0$ is the linear single-particle Schr\"{o}dinger
Hamiltonian. Furthermore, in order to reduce Maxwell's equations  to
three scalar equations, it has been assumed that $L_n \gg \lambda_L$
and 
${\boldsymbol{\nabla}} \epsilon \cdot {\bf{E}} \simeq 0$ ($L_n$ is the
characteristic length scale of transverse density modulations and
$\lambda_L$ is the radiation wavelength).
Notice that the atom equation is the approximate version of what
was presented in \cite{ref:krutitsky},  valid only  under the assumptions of a
semi--classical approach which limits the model to a well defined range of parameters: The
concept of force being purely 
classical, quantum fluctuations, stochastic heating and any
incoherent process are to be neglected which is a consistent assumption if 
large detunings are considered $|\Delta| \gg
\omega_a, \Gamma$ ($\Gamma$ is the natural line
width of the atoms). Although these limitations are
quite strict,  the model should elucidate the basic physics
of the interaction and hopefully
the resulting structures will be resilient enough to be interesting
even under non ideal conditions.\\
Finally, for a mean field model to be valid for the atom wave function, we must
consider not only the zero temperature limit but also 
a low density limit with $n a_s^3 \ll 1$, see
\cite{ref:review}. Furthermore, a low density regime is required in
order to avoid the singularity of the model and concomitant spurious
collapse-like phenomena. In view of this limitation and in order to
simplify the analysis, in what follows we will neglect the
denominator in both equations. This is a delicate step and cannot
be taken when analysing the coupled dynamics of the system, since it is
not guaranteed that during the evolution the peak values of the atom density
will satisfy the low density assumption  (as seen in
  numerical simulations of the coupled evolution, \cite{ref:us3},  the nonlinear
  focusing action of the laser could be strong enough as to focus the
  atoms to very high peak densities in a single spot thus breaking the
  low density assumption). However, it is possible to
accept this simplification when studying the stationary solutions with
the proviso that only low density solutions will be accepted.

\section{The physical setup of the problem}

In general, if laser and atoms are prepared in an initial state which
is not a fully stationary state of the system, we expect the system to
evolve showing changes in both the atom wave function and the laser
amplitude profile with the propagation variable (hereafter chosen to
be $z$).Thus it was observed in \cite{ref:us1} and \cite{ref:us2} via numerical simulations of the coupled
equations that, depending on the parameters of the initial state, the
propagation effects could have different outcomes. In particular,
starting from a super--Gaussian laser amplitude much wider than the
initial Gaussian atom density distribution, a regime of parameters
(peak laser intensity and peak atom density) could be found for which
the initial single-peak atom wave function and the flat laser
amplitude profile slowly changed into two symmetrical peaks. Once
these structures were formed, further propagation led to different
possible scenarios: (1) outward motion of the two peaks moving farther
apart from each other (this is what we call ''jet emission''); (2)
inward motion of the two peaks resulting in coalescence into a single
central peak; or (3) inward motion of the two peaks resulting into a
bound state with the two structures oscillating about the central
position. Thus, from simulations of the coupled propagation equations
it seems possible to see the creation of such symmetrical double-hump
structures which however do not survive for long and undergo their own
motion. In order to understand their nature and destiny (can they be
thought of as solitary waves? Can we describe the outcomes of their
coupled propagation in terms of soliton--soliton interactions?), we
need to know what sort of stationary structures the model admits and
work out the properties of their interactions.

\section{Stationary solutions}
The existence of stationary solutions and their stability is
fundamental from the point of view of realisable structures.
We are interested in stationary solutions corresponding to mutual
guiding in the form of long distance localised beam propagation. To
analyse these solutions we focus on modes that are localised in the
transverse direction, the analogue of Kerr spatial solitons, by
assuming
\begin{eqnarray}
{\boldsymbol{\mathcal{E}}}({\bf{r}}) = a({\bf{r}}_\perp) \exp(i h_a z)
{\bf{e}}\\
\Phi({\bf{r}}) = \phi({\bf{r}}_\perp) \exp(i h_\phi z)
\end{eqnarray}
where ${\bf r}_\perp$ denotes the dimension transverse to the
propagation direction $z$, ${\bf e}$ is the polarization vector  of
the electric field, $a, \phi$ are real amplitudes  and $h_a$, $h_\phi$ are the laser and atom propagation
constants respectively. These solutions must satisfy the fully
stationary equations derived from (\ref{eq:atom}) and
(\ref{eq:laser}), which when written in normalised variables become:   
\begin{eqnarray}\label{eq:atomstat}
\tilde{\nabla}_\perp^2 \tilde{\phi} - \beta\tilde{\phi}^3 +s\tilde{a}^2 \tilde{\phi} - 2 \mu^2 \kappa_\phi \tilde{\phi}=0,\\
\tilde{\nabla}_\perp^2 \tilde{a}
+3s \tilde{\phi}^2 \tilde{a} - 2 \kappa_a \tilde{a}=0.\label{eq:laserstat}
\end{eqnarray}
The normalisation used is: $\tilde{{\bf{r}}} = {{\bf{r}}}k_L$, for the
atom wave function $ \tilde{\phi} = \phi/\phi_*$ with
$(4\pi|\alpha|/3)\phi_*^2 = 1$, for the laser $\tilde{a} =a/a_*,$ with
$ m|\alpha| a_*^2/(2\hbar^2k_L^2) = 1$, $s = sign(\alpha)$, and
$\beta=6a_s/(k_L^2 |\alpha|)$ is directly proportional to the strength
of the collisional nonlinearity. The tilde will be dropped hereafter
unless otherwise stated. Furthermore, $\mu = k_a/k_L$,
$k_a=\sqrt{2m \omega_a/\hbar}$,
$\kappa_\phi=\left(h_\phi^2/k_a^2-1\right)/2$ and
$\kappa_a=\left(h_a^2/k_L^2-1\right)/2$. For simplicity, we will consider only one
transverse dimension, ${\bf r}_\perp=x$ and assume $\mu =
1$. To allow for mutual trapping, we will also assume $s = +1$.\\  
A further change of variables allows to rewrite these equations in terms
of the relative wave number only 
\begin{eqnarray}\label{eq:atomstat1}
\bar{\phi}''  +(\bar{a}^2  - 1)\bar{\phi}- \beta\bar{\phi}^3/3=0,\\
\bar{a}''+(\bar{\phi}^2 -\bar{\kappa})\bar{a}=0.\label{eq:laserstat1}
\end{eqnarray}
where $\bar{\kappa}=\kappa_a/\kappa_{\phi}$ is the relative wave
number, $\bar{x} = \tilde{x} \sqrt{2 \mu^2 \kappa_\phi}$, $\bar{a}^2 =
\tilde{a}^2/\left(2 \mu^2 \kappa_\phi\right)$, $\bar{\phi}^2 =  3
\tilde{\phi}^2/\left(2 \mu^2 \kappa_\phi\right)$. Again the bar will be dropped hereafter.

\begin{figure}
\includegraphics[scale=0.23]{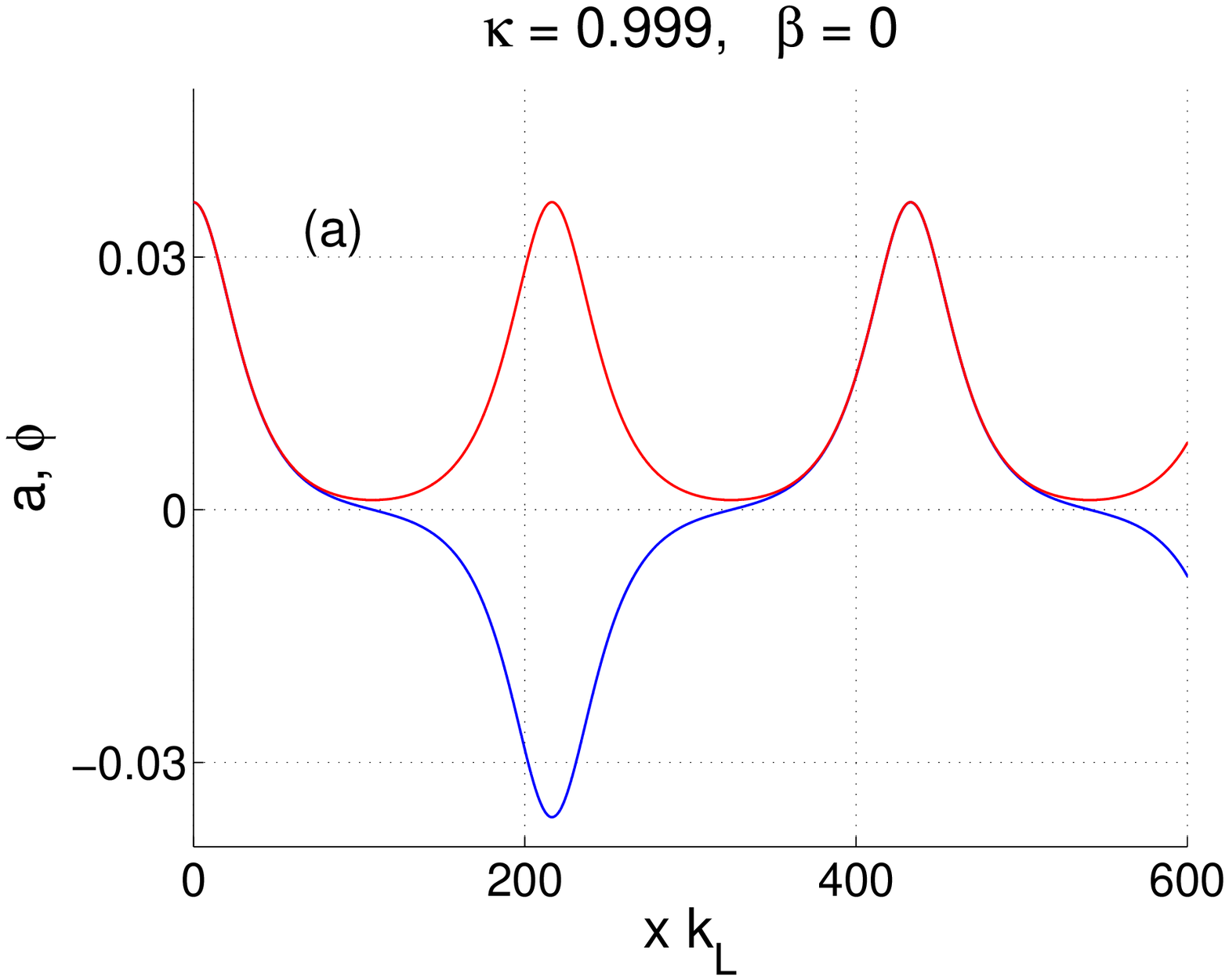}\includegraphics[scale=0.23]{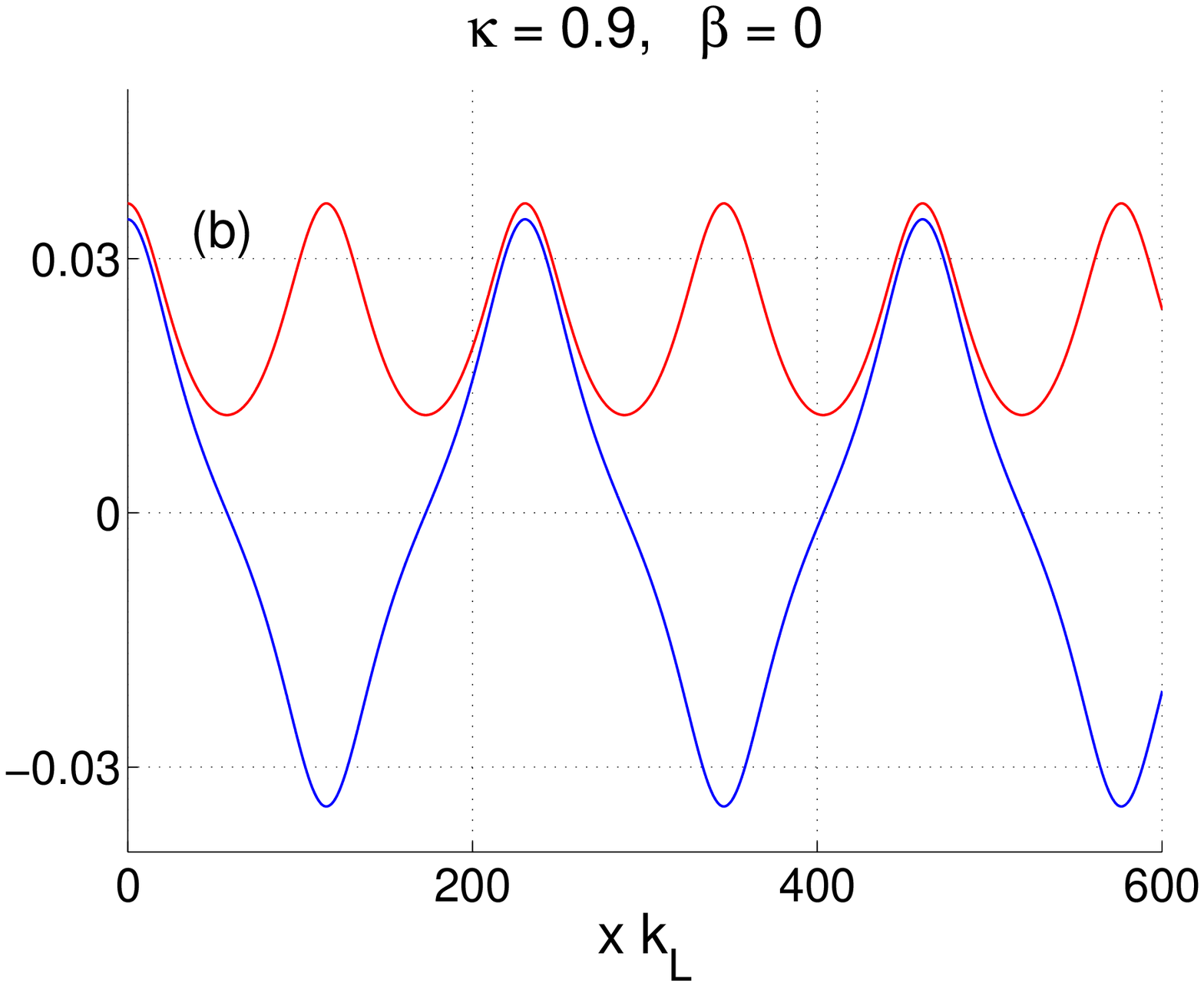}
\caption{Periodic solutions for $\beta = 0$ at different values of
  $\kappa$. Blue line $\tilde{\phi}(x)$, red line (always
  positive) $\tilde{a}(x)$. All quantities
  normalized as in the text. For $\kappa > 1$ the role of laser and
  atoms is simply inverted. Here $\kappa_\phi = 10^{-3}$. Colour online.} \label{period}
\end{figure}

To find localised solutions, we can solve this set of equations
numerically as an eigenvalue problem. 
 A first integral of motion of (\ref{eq:atomstat1}) and
(\ref{eq:laserstat1}) can be obtained by multiplying (\ref{eq:atomstat1}) by $\phi'$
and (\ref{eq:laserstat1}) by $a'$ and integrating over $x$. Combining the two resulting
integrals we find
\begin{equation}
\phi'^2+a'^2-\phi^2-\kappa a^2+\phi^2a^2-\frac{1}{6}\beta\phi^4 = \textrm{constant}, \label{hamil}
\end{equation} 
The constant is zero for a localized solution vanishing at infinity.
 From this, we can
deduce a relation between the values of $\phi$ and $a$ at  the
symmetry point $x=0$ where we require 
$\phi'(0)=a'(0) = 0$:
\begin{equation}
\phi_0^2=\frac{3}{\beta}\left(a_0^2-1-\sqrt{\left(a_0^2-1\right)^2-2\kappa
    a_0^2 \beta/3}  \right),\label{rel}
\end{equation}
where $a_0 = a(x=0), \phi_0 = \phi(x=0)$. Then $a_0$ can be used as a shooting parameter
to search for localised solutions with the shooting method
\cite{numrec}. For a fixed value of the eigenvalue $\kappa$, a value is assumed for
$a_0$ with $\phi_0$ consequently calculated from (\ref{hamil}) and a
solution is calculated with an ordinary differential equation
solver. The value of $a_0$ is than varied until a solution with the
desired characteristics is found (single--hump, double--hump \ldots).
As shown in our previous work
\cite{ref:us3}, this equation admits single hump  soliton--like solutions, in
particular, it was shown that for one-scaled distributions ($\kappa=1$) the set of
eqs.~(\ref{eq:atomstat1}), (\ref{eq:laserstat1}) reduces to one
equation if  $a(x)=\sqrt{\beta/3+1}\phi(x)$, the soliton solution
corresponding to $a_0 = \sqrt{2}$ as is evident when solving the
resulting nonlinear Schr\"{o}dinger equation. We have expanded the family of
solutions of (\ref{eq:atomstat1}), (\ref{eq:laserstat1}) and it
is now clear that this set of equations admits multi--hump and
periodic  solutions as well for which the Hamiltonian (\ref{hamil}) is exactly
equal to zero.   
This can be easily
seen in the simple case of $\beta=0$. At $\kappa \neq 1$ with $a_0=\sqrt{2}$, the
coupled equations admit periodic solutions, see for instance fig.\ref{period}   

\begin{figure}
\includegraphics[scale=0.23]{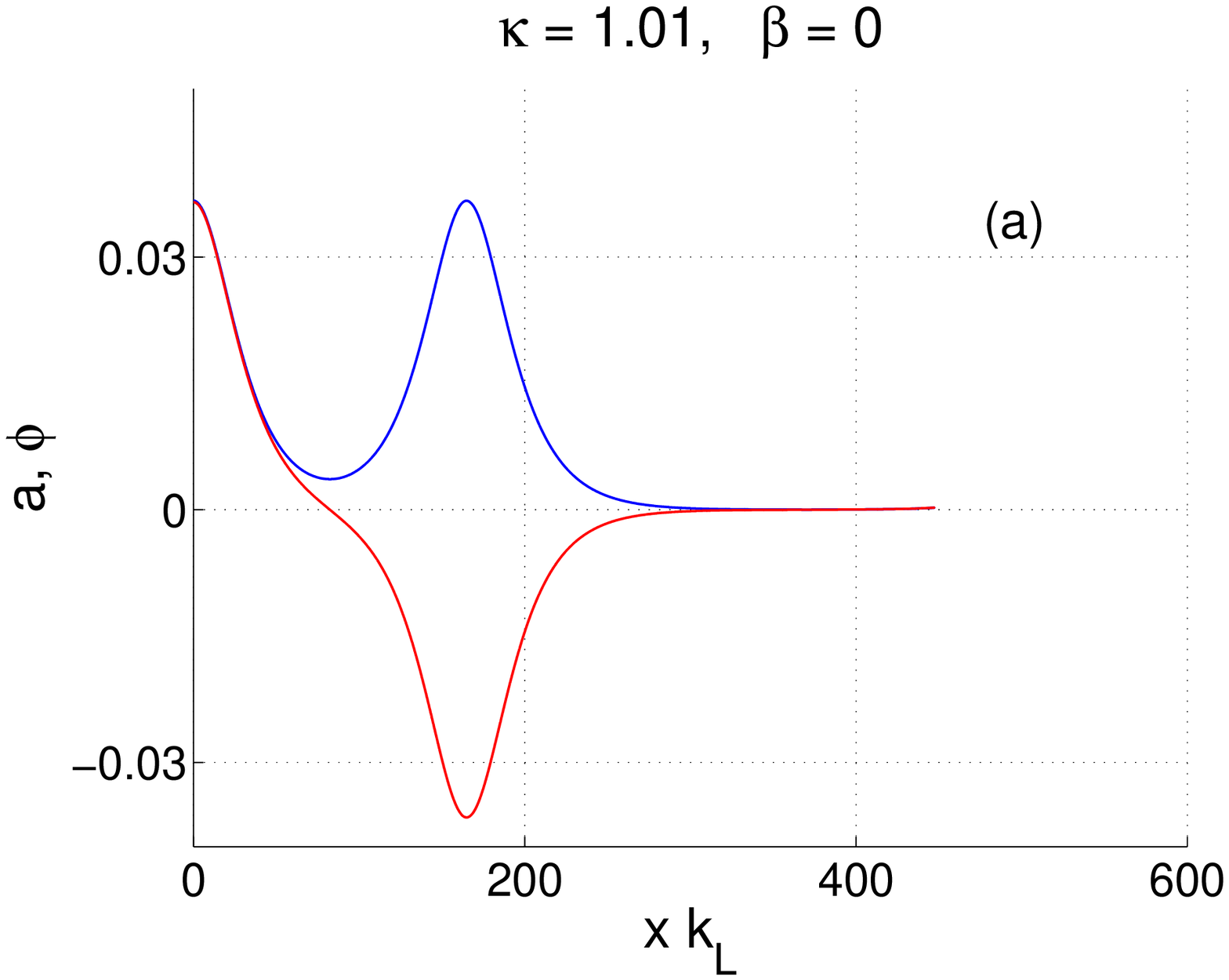}\includegraphics[scale=0.23]{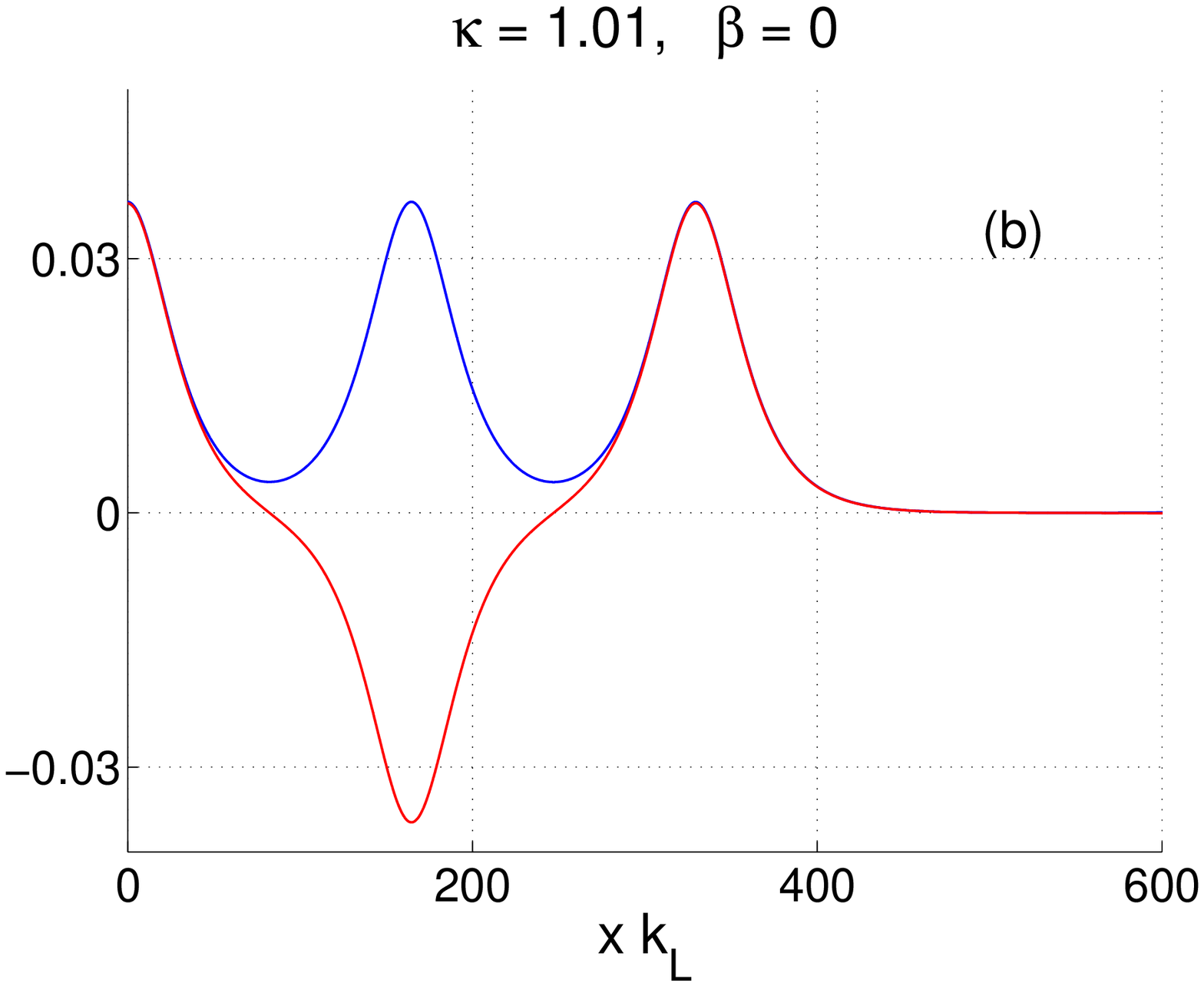}
\includegraphics[scale=0.23]{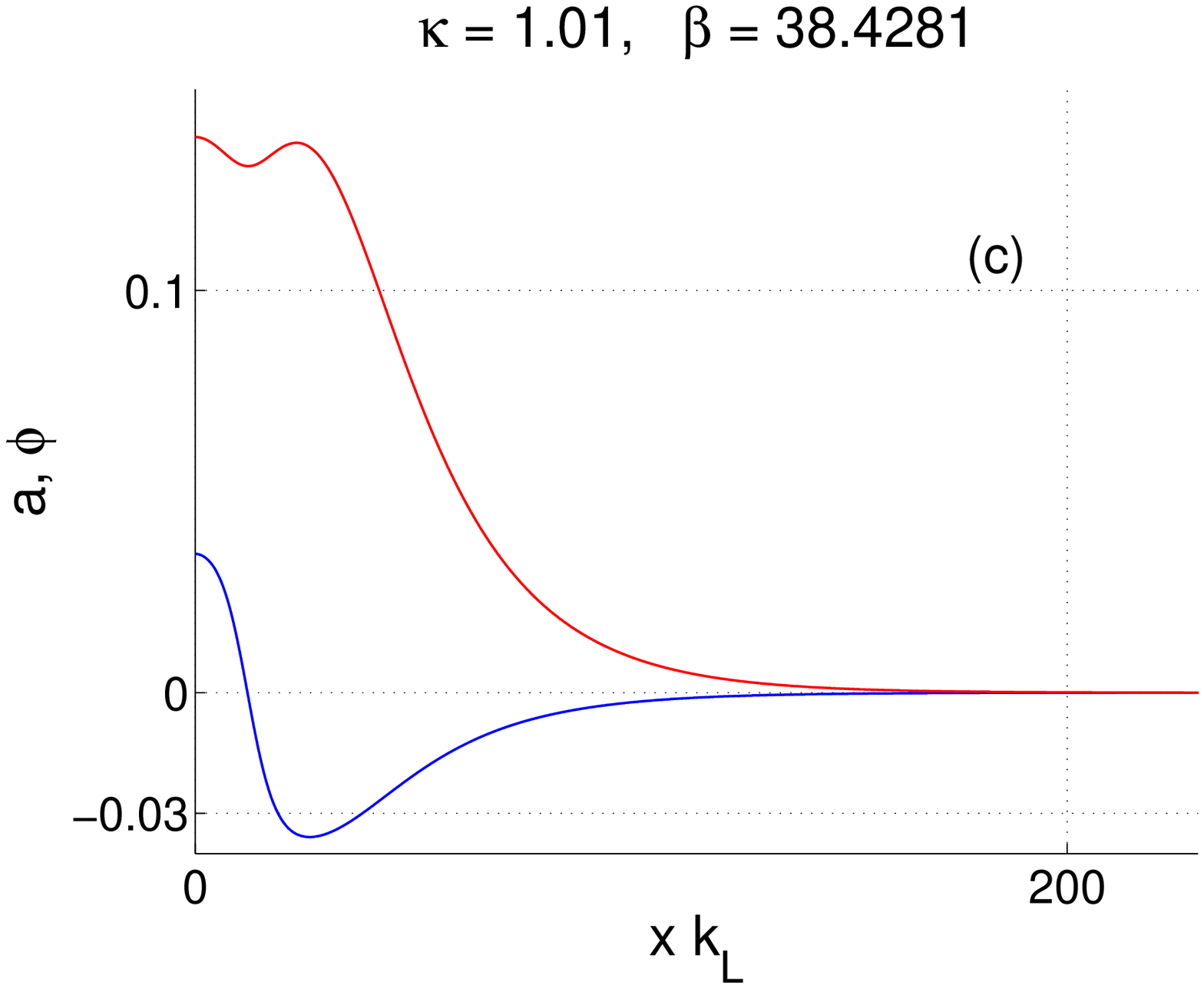}\includegraphics[scale=0.23]{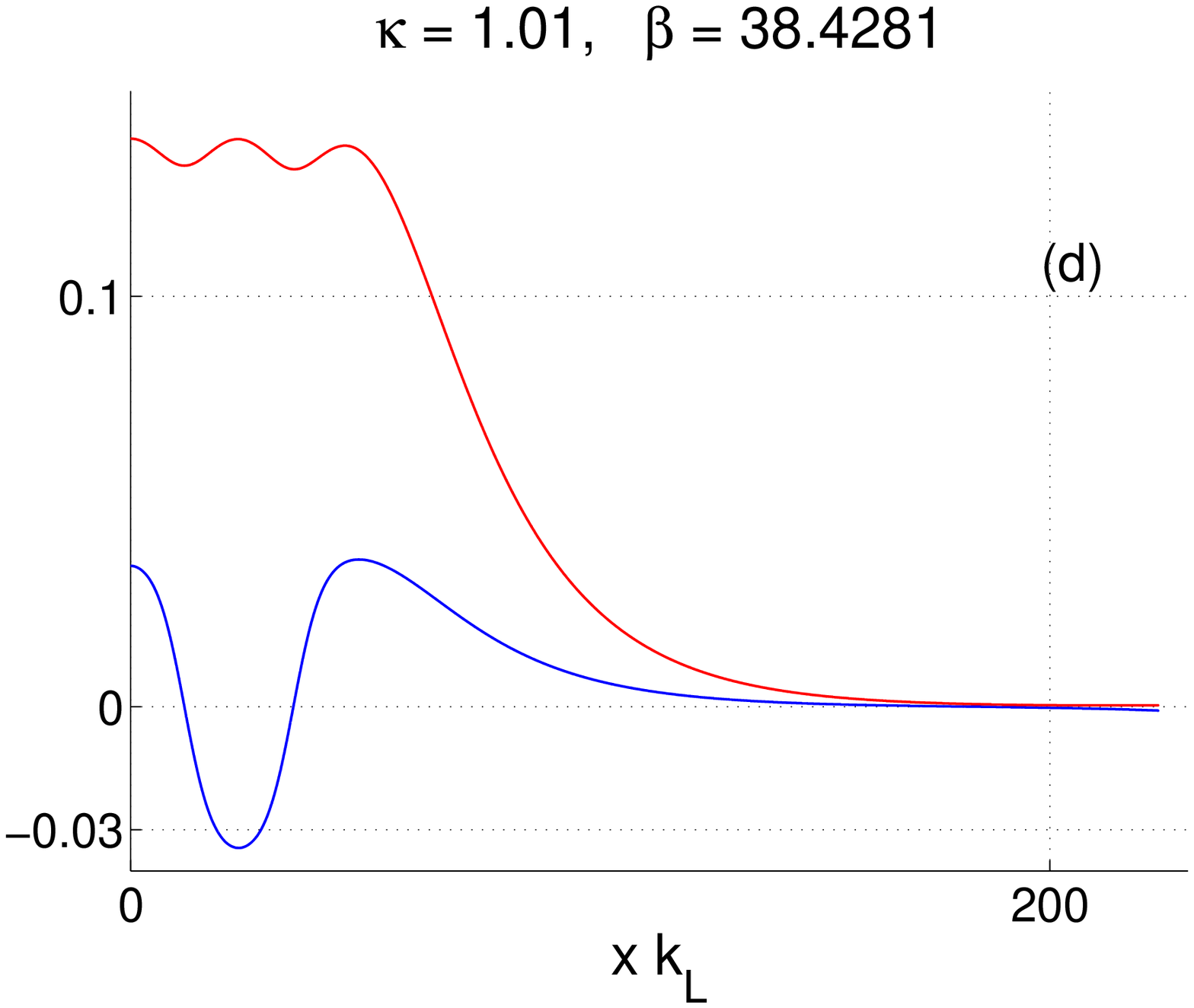}
\caption{Multi-hump solutions at $\kappa=1.01, \beta = 0$ [(a) and
  (b)] and $38.4281$ [(c) and
  (d)]. Blue line (always positive in (a) and (b))
  $\tilde{\phi}(x)$, red line (always positive in (c) and
  (d) $\tilde{a}(x)$. All quantities 
  normalized as in the text. Shooting on
the peak laser amplitude $a_0$ it is possible to identify different
solutions with increasing number of peaks. Here $\kappa_\phi = 10^{-3}$. Colour online.} \label{multihump}
\end{figure}

Multi--hump
solutions can be found as well, fixing $\kappa$ and varying $a_0$.
Some examples for $\beta = 0$ are
shown in fig.\ref{multihump}(a) and (b)  while part (c) and (d) show
examples of different  multi--hump solutions for $\beta \simeq 38$
corresponding for instance to a detuning of 100 times the decay rate
for $^{87}$Rb atoms (the same value used in the previous numerical simulations).    
By varying the shooting parameter $a_0$ it is possible to find
solutions with different number of peaks,
thus it seems indeed possible to find multi--hump solutions for any
value of $\beta$.
 
\section{Jet emission} 
We would like to underline the fact that these kind of multi--peak
  structure was never clearly seen in numerical simulations of the
  coupled propagation equation. As mentioned in the sectiom ''The
  physical setup of the problem'',  we mainly observed the generation of
  two symmetric peaks (or four at higher initial atom densities) thus
  we cannot say anything about the generation of many--peak solutions
  within the range of parameters we have explored (the main limit
  being low atom densities). It is very likely that they can be
  generated at higher atom densities but appear only as a transient to
  transform quickly into more stable structures with two peaks only.
 This has obvious  consequences for what could be oserved
 experimentally therefore it is important to unravel the dynamics of
 the structures. In particular we shall study the simplest case of two
 symmetric structures as the model case for this kind of dynamics. A
 knowledge of  how they evolve  gives the only 
 acceptable indication of what could be observed.
Therefore, once the existence of multi--hump solutions has been established, we can
proceed to study the existence of stationary solutions that originate
the ejected jets. The simplest case found in numerical simulations,
\cite{ref:us2}, pictures an initially centrally
localised bunch of atoms coupled to a localised profile of laser
intensity (a Gaussian and a super--Gaussian respectively in the
simulations), splitting into two ''jets'' which are then ejected and
start  to propagate in opposite directions together with two laser
jets. The aim now is to find stationary solutions consisting 
of only two symmetric (or antisymmetric) localised humps. If they
exist, we expect a tail--tail interaction between them because of the
nature of the equations.
In the case of well separated humps, i.e. weak overlap and
consequently weak tail interaction, a perturbative analysis completely akin to that introduced by
Yang in \cite{ref:yang} can be used to make predictions about the fate of the two
humps. The question is whether they will propagate together or whether they will repel or attract each other because of
their interaction. 

\begin{figure}
\begin{center}
\includegraphics[scale=0.28]{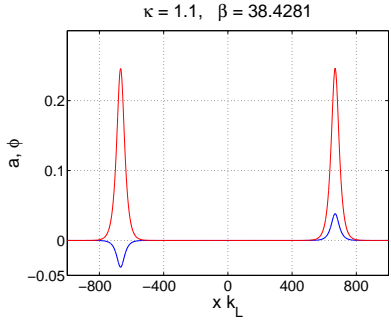}
\caption{An example of a double-hump solutions at $\kappa=1.1$ with  $\beta = 38.4281$. Blue line
  $\tilde{\phi}(x)$ (antisymmetric curve at lower
  amplitude), red line $\tilde{a}(x)$ (symmetric curve at
  higher amplitude). All quantities 
  normalized as in the text.  Here $\kappa_\phi = 10^{-3}$. Colour online.} \label{twohump}
\end{center}
\end{figure}

An example of a symmetric two-peak solution is shown in
fig.\ref{twohump} for fixed $\beta$ and $\kappa$. These structures are
found by imposing $\phi(0) = a'(0) = 0$, using   the
Hamiltonian (\ref{hamil}) to obtain an expression for $a(0)^2 = \phi'(0)^2/\kappa$
and  shooting on the value of the first derivative $\phi'(0)$.
Due to the symmetry
of the equation, it is enough to calculate one half of the solution,
say for $x>0$ and any combination (laser symmetric--atoms antisymmetric,
laser antisymmetric--atoms symmetric and so on) is still a solution.

\begin{figure}
\begin{center}
\includegraphics[scale=0.28]{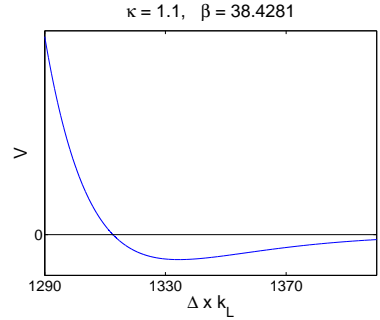}
\caption{Yang's potential $V$ versus the soliton separation $\Delta
  \tilde{x}$ for  $\kappa=1.1$ and  $\beta = 38.4281$. } \label{potential}
\end{center}
\end{figure}

With a known double--peak solution showing two very well separated
peaks, it is possible to study the nature of the
interaction between the tails of the two peaks. We have followed the procedure
by Yang in \cite{ref:yang}, who introduced a perturbative
analysis of the interaction of vector solitons which we shall only
briefly recall here. Modifying the
perturbative approach of Karpman and Solov'ev, Yang found dynamical
equations for the parameters of two--vector solitons weakly interacting
with each other because widely separated. In particular, for cases in
which the two components have phase difference $0$ or $\pi$, Yang determined
an equation for the dynamics of the vector--soliton separation $\Delta
x$ in the
form of a Newtonian equation of motion $m \, d^2 \Delta x/ dz^2 = -
\nabla V$. The fixed points of this equation give the separation distance
of the two components of a stationary solution. We can
  therefore apply Yang's formulation to the main case under analysis,
  that of a two--peak stationary solution as the one numerically
  evaluated and shown in fig.\ref{twohump}. This
  solution corresponds to what seen in numerical simulations of the
  atom--laser coupled propagation (see the section ''The physical setup of
  the problem'') before the two peaks start to move either inward or
  outward. With a detuning of about 100 times the decay rate for
  $^{87}$Rb atoms (i.e., $\beta \simeq 38$), we find a peak atom
  density of $2.49 \,10^{19} $ m$^{-3}$ and a peak laser intensity of
  $0.051 $ mW/cm$^2$.
Utilizing Yang's
formula, we arrive for the case shown in fig.\ref{twohump} at
$\Delta \tilde{x}_0 = 1.3344 \, 10^3$ which compares quite well with the
separation found numerically $1.3359 \, 10^3$. Therefore we can use
the potential $V$ to infer the dynamics of the separation $\Delta
x$. Again for the case shown in fig.\ref{twohump} and using Yang's
formula for $V$, the potential is shown in fig.\ref{potential}. It is
clear that if the two components are generated at a distance $\Delta
\tilde{x} < \Delta \tilde{x}_0$, the two solitons will repel each
other and be ejected away whereas two components generated at $\Delta
\tilde{x} > \Delta \tilde{x}_0$ will attract each other with the
possible creation of a bound state where the two components oscillate
about their equilibrium position. This model,
based on the existence of multi--peaks solutions of our stationary equations, 
thus explains quite well the qualitative features of 
the numerical results of \cite{ref:us2}, both the jet emission and the
formation of a bound state. It thus seems possible to say that the dipole--dipole interaction can
  lead to the generation of two--peak structures for both atoms and
  laser but the destiny of the two peaks is that of moving apart from
  each other in a solitary--like fashion or to move towards each
  other. This is indeed what seen via numerical simulations of the
  coupled propagation and we interpret the two--peak solutions as the
  unstable seeds of the evolution predicted by Yang's model and
  observed numerically.

\section{Conclusions}
Numerical simulations of copropagating laser--BEC systems seem to
indicate that not only mutually localised structures can be formed, but
also that solitary--like wave packets can be emitted out of the
central interaction region. In order to understand the physics of the
jet ejection, we have used a simplified system of
coupled equations and studied the existence of possible multi--peaked
stationary solutions, usually not possible for the nonlinear
Schr\"{o}dinger equation.  Applying the results of a  perturbative
method  elaborated in nonlinear optics to study their dynamics, we
have found that jet emission and bound state formation  can be
explained by the tail interaction of the soliton components of a 
two-peak structure within the model we are using.

\end{document}